\documentclass[a4paper,11pt]{article}

\pdfoutput=1 
             
\usepackage{graphicx} 
\usepackage{subcaption} 

\usepackage{jcappub} 
\usepackage{lineno}

\usepackage[T1]{fontenc} 
\usepackage{rotating} 

\usepackage{cancel}

\usepackage{amsthm}

\theoremstyle{thmstyleone}%
\theoremstyle{thmstyletwo}%

\theoremstyle{thmstylethree}%
\newcommand{\refF}[1]{Fig.~\ref{#1}}

\newcommand{\refS}[1]{section~\ref{#1}}
\newcommand{\refT}[1]{table~\ref{#1}}
\newcommand{\refApp}[1]{appendix~\ref{#1}}

\usepackage{orcidlink}
\usepackage{hyperref}

\title{Simple $\phi\Lambda$CDM dynamics \\ Accepted: 27 Feb 2025}

\author[a]{Pierros Ntelis \orcidlink{0000-0002-7849-2418}}
\author[b,c]{,Jackson Levi Said \orcidlink{0000-0002-7835-4365}}

\affiliation[a]{Independent Research Affiliation, Marseille, France}
\affiliation[b]{Institute of Space Sciences and Astronomy, University of Malta, Msida, Malta}
\affiliation[c]{Department of Physics, University of Malta, Msida, Malta}

\emailAdd{ntelis.pierros <at> gmail <point> com}
\emailAdd{jackson.said@um.edu.mt}

\abstract{
In this paper, we introduce the $\phi\Lambda$CDM model and compare it to the concordance model. We present their respective epoch evolutions, perform a detailed dynamical analysis for each model, and conduct a comparative analysis between the two. This study revitalizes these models by considering systems with a higher number of variables. Additionally, the $\phi\Lambda$CDM model we present is both more comprehensive and simpler than those found in the literature, as it accounts for all known epochs, including the radiation, matter, and dark energy phases. Notably, existing studies on this model often omit the radiation epoch and focus on simpler dynamics.

We find that both models, the $\phi\Lambda$CDM and the $\Lambda$CDM,  can describe the generally accepted scenario of cosmic evolution, and current observations. Both models, describe qualitative and quantitative current observations about the epoch behaviour of the species of the universe. We find the $\phi\Lambda$CDM model has the following exotic transition 
from a 
\textit{dominant radiation energy density ratio} 
epoch, 
or a \textit{low scalar kinetic term energy density ratio} 
epoch, 
towards a 
\textit{dominant cosmological constant energy density ratio}
epoch. This renders the $\phi\Lambda$CDM a richer phenomenologically model than the concordance cosmological model.

The software of the study is publicly available \href{https://github.com/lontelis/Simple_quintessence_dynamics}{\texttt{SimplePhiLCDM}}.
}


\begin{document}
\maketitle
\date
\flushbottom

\section{Introduction}\label{Introduction}

Currently, most extensions of gravity are constructed by modifying the components of the action. Typically, these modifications involve altering either the geometric aspects or the matter Lagrangian, with the goal of addressing and potentially resolving cosmological tensions \cite{Schoneberg:2021qvd} alongside other challenges associated with the concordance model. In this work, we assume a homogeneous and isotropic universe \cite{Ntelis:2016suu,Laurent,Ntelis:2016utg,2017JCAP...06..019N,Ntelis:2018tlj}. Our approach aims to address cosmological tensions \cite{Schoneberg:2021qvd} by focusing on Riemannian geometries with the inclusion of additional scalar fields, an area that has garnered significant interest recently \cite{CANTATA:2021ktz}.

The research community has extensively explored modifications of the geometric aspects of gravitational theory \cite{CANTATA:2021ktz,Bahamonde:2017ize,Bahamonde:2020lsm,Bahamonde:2021akc,Aoki:2023sum,Bahamonde:2021gfp}; however, such modifications are not considered in this work. Notably, there are various alternative extensions to gravity that involve constructing mathematical models using the action as a fundamental component, such as $f(R)$ theories, Horndeski gravity, non-Riemannian cosmologies, and functor of actions theories \cite{2023FoPh...53...29N,Ntelis:2024fzh}. Nevertheless, these extensions are also beyond the scope of our current analysis. To simplify our approach and avoid potential complications, we focus on introducing an additional species as a new ingredient to construct the $\phi\Lambda$CDM model. This ingredient is a scalar dynamical field, denoted by $\phi(t)$, which allows us to build a model of dynamical dark energy. In this framework, the dark energy component is characterized not only by the evolution of the scalar field $\phi(t)$ but also by its potential 
$V[t,\Lambda;\phi(t)]$, which explicitly depends on both the scalar field and the cosmological constant $\Lambda$.

The $\phi\Lambda$CDM model has been previously studied in the literature \cite{Copeland:1997et,Copeland2006,vandeBruck:2016jgg,Ramadan:2023ivw}. Our study differs from that of \citet{Copeland:1997et,Copeland2006}, who consider a single barotropic fluid model, by instead distinguishing between the fluids of matter and radiation separately within the model. While \citet{vandeBruck:2016jgg} focus on an interacting dark energy model and include some discussion of $\phi\Lambda$CDM dynamics, our analysis is centered exclusively on the $\phi\Lambda$CDM model, offering additional details in the dynamical analysis. In our approach, we separately consider the total matter energy density, the kinetic energy density of the scalar field, and the potential energy density of the scalar field. This differentiates our model from that of \citet{Ramadan:2023ivw}, who also include three fluids but focus on the total radiation energy density rather than the matter energy density. Moreover, our study extends beyond previous analyses by providing a comprehensive examination of the $\phi\Lambda$CDM model. Specifically, we present all critical points, detailed numerical solutions, and an in-depth exploration of the scalar field potential configuration within the dynamical system.

In contrast with the previous studies, we develop a dynamical analysis \cite{Bahamonde:2017ize} on both $\Lambda$CDM and $\phi\Lambda$CDM, with 3 variables, constructing 3D systems of the models, 
we solve them analytically and numerically, and provide their phase portraits through the dynamical and stability analysis, and we apply a comparative analysis, with many more interpretations. 

In this work, we analyze the dynamical properties of cosmological models using a phase-space approach based on dimensionless energy density ratios. For the \(\Lambda\)CDM model, we consider the variables \((\Omega_m, \Omega_r, \Omega_\Lambda)\), corresponding to the fractional energy densities of matter, radiation, and the cosmological constant, respectively. This formulation captures the evolution of the dominant energy components without requiring explicit dependence on additional cosmological parameters, such as the Hubble constant or baryon density. 
For \(\phi\Lambda\)CDM model incorporating a scalar field, we adopt a similar framework, replacing \(\Omega_\Lambda\) with \(\Omega_x\) and \(\Omega_v\), which represent the kinetic and potential energy density contributions of the scalar field. This choice allows for a direct comparison between \(\Lambda\)CDM and modified gravity scenarios while maintaining a consistent dynamical system formulation. 
By reducing the analysis to a minimal set of variables that govern the qualitative evolution of the universe, we focus on the stability and critical points of the system, providing insights into the late-time behavior of these models.

A notable difference with previous studies is that  we consider separately the continuity equation for the kinetic term, and the potential term, we end up to the system which is simpler than the previous studies, as done recently \cite{Ntelis_Said:2024}.  
This means that we do not need to define extra variables, such as the $\lambda$, and $\Gamma$, as in previous studies. Note that \(\lambda\) is introduced to characterize the shape of the exponential potential and is related to the first derivative of the potential with respect to the scalar field. Similarly, \(\Gamma\) is associated with the second derivative of the potential. However, since our analysis does not explicitly involve these derivatives, these parameters are not relevant in our formulation.

Our focus encompasses the physics governing the background evolution of the universe during both early and late epochs. At early times, this involves exploring initial conditions and inflationary dynamics, while at late times, it addresses the accelerated expansion and the formation of large-scale structures. To address existing cosmological tensions, we will simulate equations governing motions, energy densities, and the characteristic scales of the universe, with the aim of establishing a foundation for investigating non-Riemannian cosmologies. The primary observables include ratios of matter, radiation, and dark energy (both simple and dynamical) densities, as well as the Hubble expansion rate. By reanalyzing data from current surveys, such as Euclid \cite{EUCLID:2020jwq,Euclid:2021qvm,Euclid:2024yrr} and DESI \cite{Alam:2020jdv}, using these novel models, we aim to gain deeper insights into the robustness of the cosmological model.

Through these efforts, we aim to provide a more comprehensive model of the universe, addressing existing tensions and contributing to the broader field of cosmology. By integrating the Riemannian geometry, a scalar dynamical field and its potential into cosmological models, we hope to offer novel explanations for observed discrepancies and enhance our understanding of the universe's fundamental properties.

This paper is structured as follows: 
In \refS{sec:overall_methodology}, we discuss the overall methodology, and  we present physical preliminaries for the study.
In \refS{sec:Dynamical_Analysis_phiLambdaCDM}, we describe the dynamical analysis of the $\phi\Lambda$CDM model.
In \refS{sec:dimensionless_variables_and_representative_3DphiLambdaCDM}, we present the numerical solutions of $\phi\Lambda$CDM.
In \refS{sec:phase_portraits_of_the_6D_to_3D_system}, we present the $\phi\Lambda$CDM phase portraits.
In we compare the numerical solutions between $\phi\Lambda$CDM and $\phi\Lambda$CDM,
while, in \refS{sec:phasespace_criticalpoints_comparison_of_models}, we discuss the results of all $\phi\Lambda$CDM phase portraits and compare them to the $\Lambda$CDM one. Finally, in \refF{sec:conclusion_discussion}, we conclude and discuss our results.

\section{Overall methodology and physical preliminaries}\label{sec:overall_methodology}

\paragraph{}The methodology involves developing new theoretical models based on a Riemannian geometries and an additional scalar field. The methodology can be detailed as follows.
We derive the equations of motion of the $\Lambda$CDM and $\phi\Lambda$CDM models.
We provide analytical and numerical solutions to these equations, which will provide us with the actual evolution of the key observables. 
We apply a dynamical system analysis, marking appropriately the key observables. We present the results of the dynamical analysis through phase portraits.
We provide a comparative analysis between the two models by comparing their numerical solutions
We provide a comparative analysis of their phase portraits.

The standard model of cosmology aims to describe the universe as a whole. Therefore we need to define our time variables, and their current observational values of these epochs. The epoch variables for the cosmic time, $t$, which use in our analysis is  the lapse function is related to the Hubble expansion rate as $N = \ln a(t) $, which depends on the scale factor, $a(t)$. 

We assume the following species that fill the universe: the total matter of the universe, $m$, which contains the baryonic, lepton and the cold dark matter; the radiation of the universe, $r$, which contains the total amount of photons and massless neutrinos of the universe; the dark energy, which is modelled either by just the cosmological constant, $\Lambda$, or both the cosmological constant, and the dynamical scalar field, $\phi(t)$. Therefore we write that the species variable takes the values:
	$ s \in \{ m, r, \Lambda, \phi\} \; .$ We describe the energy density ratio of each species with, $\Omega_s=\Omega_s[t;N(t)]$.

Initially, at some epoch of initial time, $t_{i}$, we can assume that
\( N_i = -12 \).
Then this means that the scale factor at that initially we have the following lapse function redshift, and scale factor triplet:
\(
	(N_i, z_i, a(t_i) ) \simeq (-12, 1.6 \times 10^{5}, 6 \times 10^{-6}) \; .
\)
Note the this assumption, under the standard $\Lambda$CDM model corresponds to an initial time of the universe, which corresponds to a cosmic lookback time, $t_{\rm clb}=13.46$ Billion years. Therefore, this initial value for the lapse function is a good assumption.


\section{Dynamical analysis on $\phi\Lambda$CDM}\label{sec:Dynamical_Analysis_phiLambdaCDM}

In this section, we describe the $\phi\Lambda$CDM, and then we describe the dynamical analysis (DA) applied to it. The $\phi\Lambda$CDM model is built on an action which has minimally coupled dynamical scalar potential, $\phi$, is a scalar dynamical field, with a kinetic term, described by partial derivatives, $\partial \phi$, and $V=V[t, \Lambda; \phi(t)]$ is its potential. Then by applying the least action principle, $\delta S_{\phi\Lambda\text{CDM}}=0$, we get the following set of equations.
The $\phi\Lambda$CDM Friedmann equations are 
\begin{eqnarray}
	\label{eq:Friedman_1st_compact_phiLambdaCDM}
	3H^2(t) &= \kappa^2 \sum_{s \in \left\{ m, r, \phi \right\}} \bar{\rho}_{\rm s}(t) \\
	\label{eq:Friedman_2nd_compact_phiLambdaCDM}	
	2\dot{H}(t) + 3H^2(t) &= - \kappa^2\sum_{s \in \left\{ m, r, \phi \right\}} w_s(t) \bar{\rho}_{\rm s}(t) 
\end{eqnarray}
where $H(t)$ is the hubble expansion rate, $s$ is index of the species, in our case, $m$ matter, $r$ radiation, $\phi$ dynamical scalar field, where $\kappa^2 = 8\pi G_{\rm N}/c^2$, where $G_{\rm N}$ is the Newton gravitational constant, and $c$ the speed of light.
The Klein-Gordon (KG) equation for the dynamical scalar field, $\phi$, is written as
\begin{eqnarray}
	\ddot{\phi} + 3H \dot{\phi} +  V_{, \phi} = 0
\end{eqnarray}
where $V_{, \phi}=\partial_\phi V(\phi)=\frac{dV(\phi)}{d\phi}$ is the derivative of the potential, in respect of the dynamical scalar field. 
Note that the dynamical system is independent of the choice of the potential, or the shape of the actual scalar field. Therefore, the KG equation is omitted for the DA. This is distinct difference from previous studies.
In \refApp{sec:description_of_Lambda_in_phiLambdaCDM_model}, we describe how the $\Lambda$ appears in $\phi\Lambda$CDM model, and how this model reduces to $\Lambda$CDM model. 

\subsection{Dimensionless variables and the representative 3D system}\label{sec:dimensionless_variables_and_representative_3DphiLambdaCDM}

Assuming that the equations of states are well known for these fields, and fixed to the following values, $\left\{ w_m, w_r \right\} = 	\left\{ 0, 1/3\right\}$, while the scalar-field equation of state, $w_\phi = w_{\phi} (x,v)= \frac{x - v}{x + v}$.

We define the following dimensionless variables
\begin{align}
 m= \Omega_{\rm m} &= \frac{\kappa^2 \bar{\rho}_m(t)}{3 H^2(t)} \quad , \quad r = \Omega_{\rm r} =  \frac{\kappa^2 \bar{\rho}_r(t)}{3 H^2(t)} \\
  x = \Omega_{\rm x} = \Omega_{\rm kin} 
&= \frac{\kappa^2}{3 H^2}X =  \frac{\kappa^2}{3 H^2} \frac{\dot{\phi}^2}{2} =  \frac{\kappa^2 \dot{\phi}^2 }{6 H^2}   \\ 
  v = \Omega_{\rm v} = \Omega_{\rm pot} 
&= \frac{\kappa^2}{3 H^2}V[t;\phi(t),\Lambda]
\end{align}

Note that the effective equation of state is written as 
\begin{eqnarray}
	w_{\rm eff} 
	&= \sum_{s \in \left\{ m,r, x,v\right\}} w_s(t) \Omega_s(t) = 
	\frac{1}{3} r(t) + x(t) - v(t)
\end{eqnarray}

With a specific choice of the parameters of the modelled guided by the choice of the potential, we end up with a representative 3D $\phi\Lambda$CDM model, since we use the first Friedman equation to define the radiation evergy density ration as a function of the rest of the energy density ratios. 

Note that by considering separately the continuity equation for the kinetic term, and the potential term, we end up to the system

so the modified Friedman equations become :

\begin{eqnarray}
	m' &=&  m (1 - m + 2 x  - 4 v)     \\	
	x' &=& x (1 - m + 2 x  - 4 v)\\	
	v' &=&  v \; (4 - m + 2 x  - 4 v)
\end{eqnarray}
where the \( r=1-m-x-v\) . 

This means that we do not need to define extra variables, such as the $\lambda$, and $\Gamma$, as previous studies, have done.  Note that$\lambda$ is introduced to quantify the shape of an exponential model of the potential of the model, and therefore ti is related to the 1st derivative of the potential in respect of the scalar field, while the $\Gamma$ is related to the 2nd derivative of the potential in respect of the scalar field. Since in our analysis the 1st and 2nd derivatives of the potential do not appear, these parameters are obsolete.

To find the critical points, we make the assumption that 
$
    (
	m',
	x',		
	v' ) 
    = \vec{0} \; .$

We solve numerical the 3D system which describes the following variables, \( 
(\Omega_m,x,v) \; .
\)

\begin{figure*}[h!]
    \centering 
    \includegraphics[width=140mm]{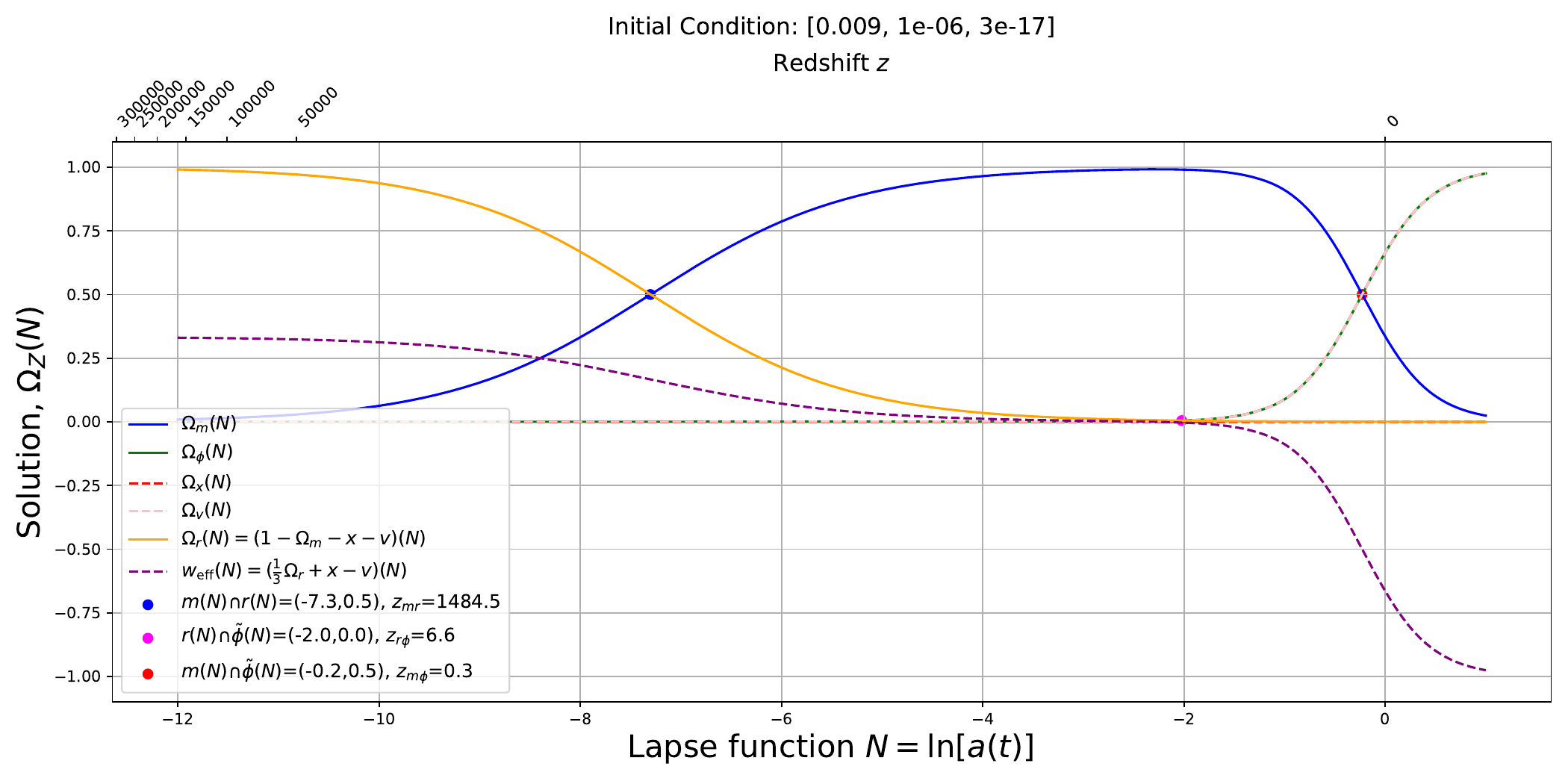}    \caption{\label{fig:ChatGPT_scalarLCDM_7D_3D_set_of_sim_diff_eqns_num_solutions_mphir_weos_with_redshift_intersections_delta_N_Nmin-12_Nmax1_zero_is_3e-17}   We illustrate the numerical solutions of the model of $\phi\Lambda$CDM cosmology, in the case which $\texttt{var-zero-is}=3 \times 10^{-17}$, and we provide more details. [See \refS{sec:dimensionless_variables_and_representative_3DphiLambdaCDM}] }
    \end{figure*}

We use the \texttt{scipy.integrate.solveivp} to solve the system of 3D system of differential equations.
We find the following numerical solutions and we present them in the \refF{fig:ChatGPT_scalarLCDM_7D_3D_set_of_sim_diff_eqns_num_solutions_mphir_weos_with_redshift_intersections_delta_N_Nmin-12_Nmax1_zero_is_3e-17}, for several specified conditions. We find the $\phi\Lambda$CDM results are similar to the $\Lambda$CDM model, on the behaviour of the epoch evolution of the different species of the model. We explain their numerical differences of the two models in \refS{sec:numerical_solutions_comparison_of_models}. What changes is the gradient  behaviour of dark energy as expected.

In particular, the potential kinetic term starts significantly low, but it is lead to a dominant scalar kinetic potential epoch in the fart future, which signifies a de Sitter universe, i.e. a universe filled with the cosmological constant. The radiation is dominant in the past, and decays as time passes. The matter starts significatly low, has a peak, during intermediate times, and then decays towards the today's epoch, and the far future. We observe mild increase of the scalar kinetic term energy density, in the late universe.

\subsection{Critical points characterisation summary}
We perform a dynamical analysis of the 3D system which describes the following variables, \(  \{ m, x, v \} \). By solving the linearised 3D system and by identifying the eigenvalues of the Jacobian of the system, and where appropriate we apply a 
Nonlinear Stability Analysis (Lyapunov's Method), defining an appropriate Lyaponov's function, $L(m,x,v)$, we find and characterise the critical points of the system. We provide the 3D points and their characterisation in \refT{tab:phiLambdaDCM_3D_points_characterised}.

\begin{table}[h!]
\centering
\resizebox{\textwidth}{!}{
\begin{tabular}{ |c|c|c|c|c|c|c|c| } 
 \hline
Point & $\Omega_m$ & $\Omega_r$ & $x$ & $v$ & $\partial_N L(m,x,v) $ & characterisation & $w_\phi$ \\ 
 \hline
 \hline
 \hline 
$R_r$ & $0$ & $1$ & $0$ & $0$ &  $>0$ & unstable, radiation domination & $0$\\ 
  \hline  
$R_m$ & $1$ & $0$ & $0$ & $0$ &  $>0$ & unstable, matter domination & $0$\\ 
  \hline      
$A_v$ & $0$ & $0$ & $0$ & $1$ &  $<0$ & stable, scalar-potential domination & $-1$\\ 
  \hline      
\end{tabular}
}
\caption{\label{tab:phiLambdaDCM_3D_points_characterised} Characterisation of critical points, using the non-linear dynamics method.}
\end{table}

\subsection{Phase portraits of the projections of 3D system}\label{sec:phase_portraits_of_the_6D_to_3D_system}
 In this section, we describe the phase spaces of the 3D system projected to the 2D spaces. Each point of the 2D diagram is related to a critical 3D point characterised in the table. 
These conditions are summarised in \refT{tab:phiLambdaDCM_3D_points_characterised}.

\begin{figure*}[h!]
    \centering
    \hspace{-2.5cm}
    \begin{subfigure}[b]{0.3\textwidth}
        \centering
        \includegraphics[width=70mm]{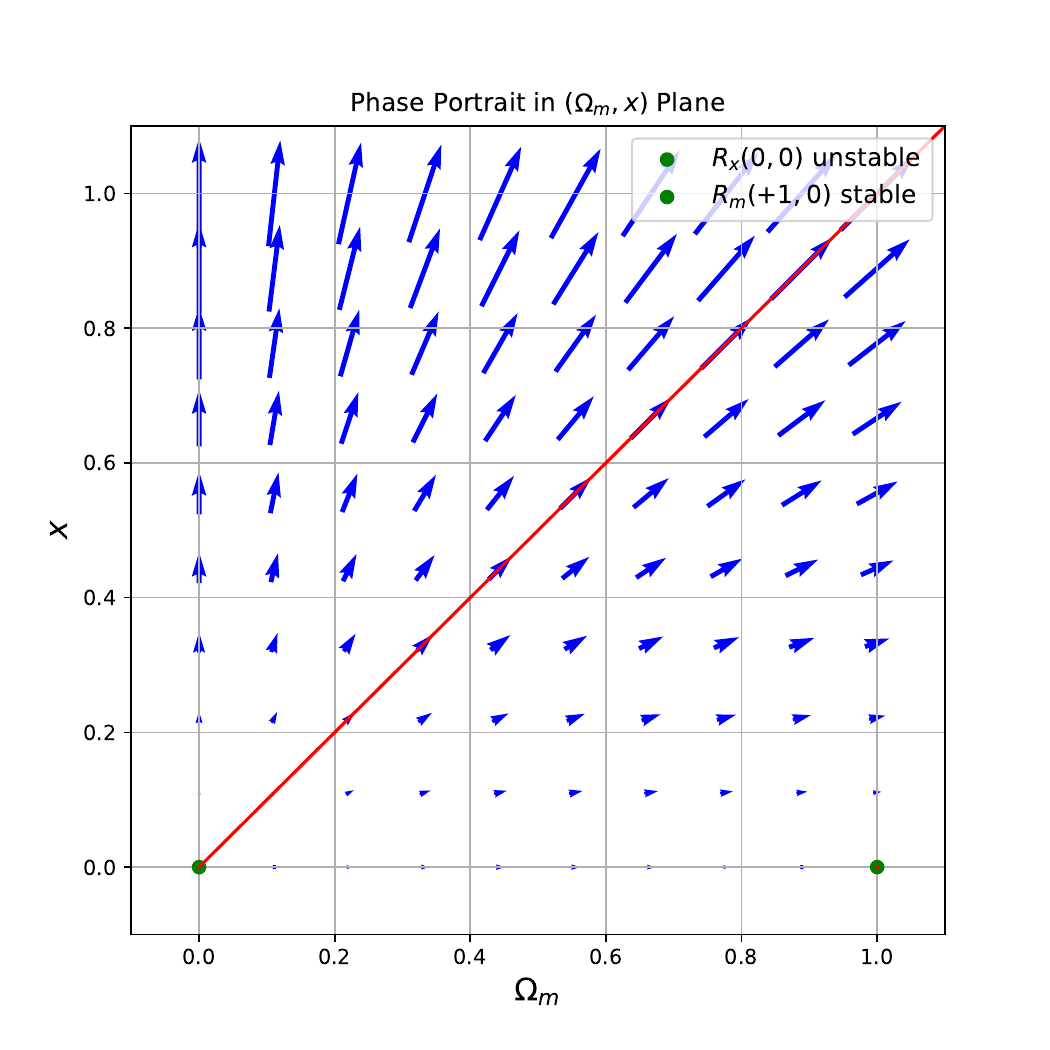}
        \label{fig:ChatGPT_scalarLCDM_7D_3D_set_of_sim_diff_eqns_phase_portraits_mx1}
    \end{subfigure}
    \hfill
    \begin{subfigure}[b]{0.3\textwidth}
        \centering
        \includegraphics[width=70mm]{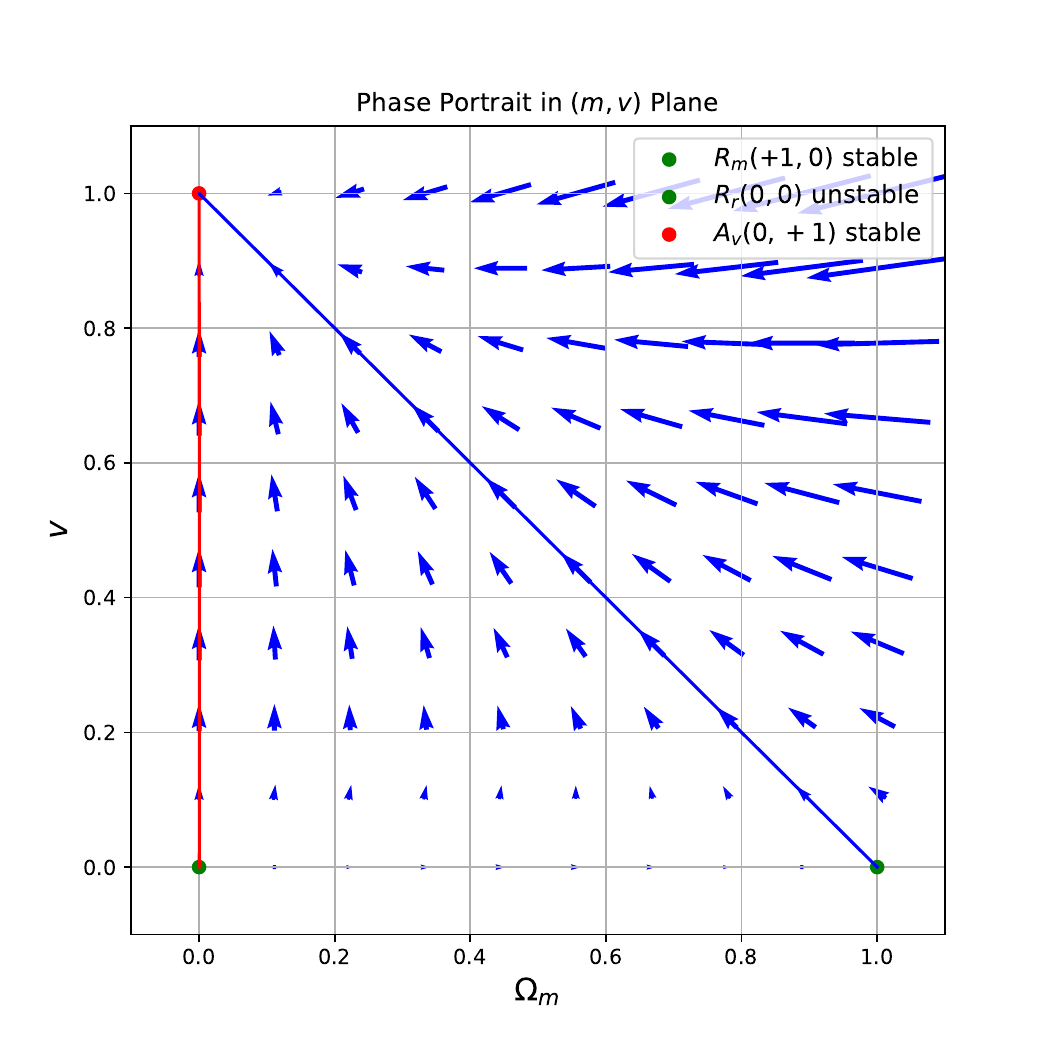}
        \label{fig:ChatGPT_scalarLCDM_7D_3D_set_of_sim_diff_eqns_phase_portraits_my1}
    \end{subfigure}
    \hfill
    \begin{subfigure}[b]{0.3\textwidth}
        \centering
        \includegraphics[width=70mm]{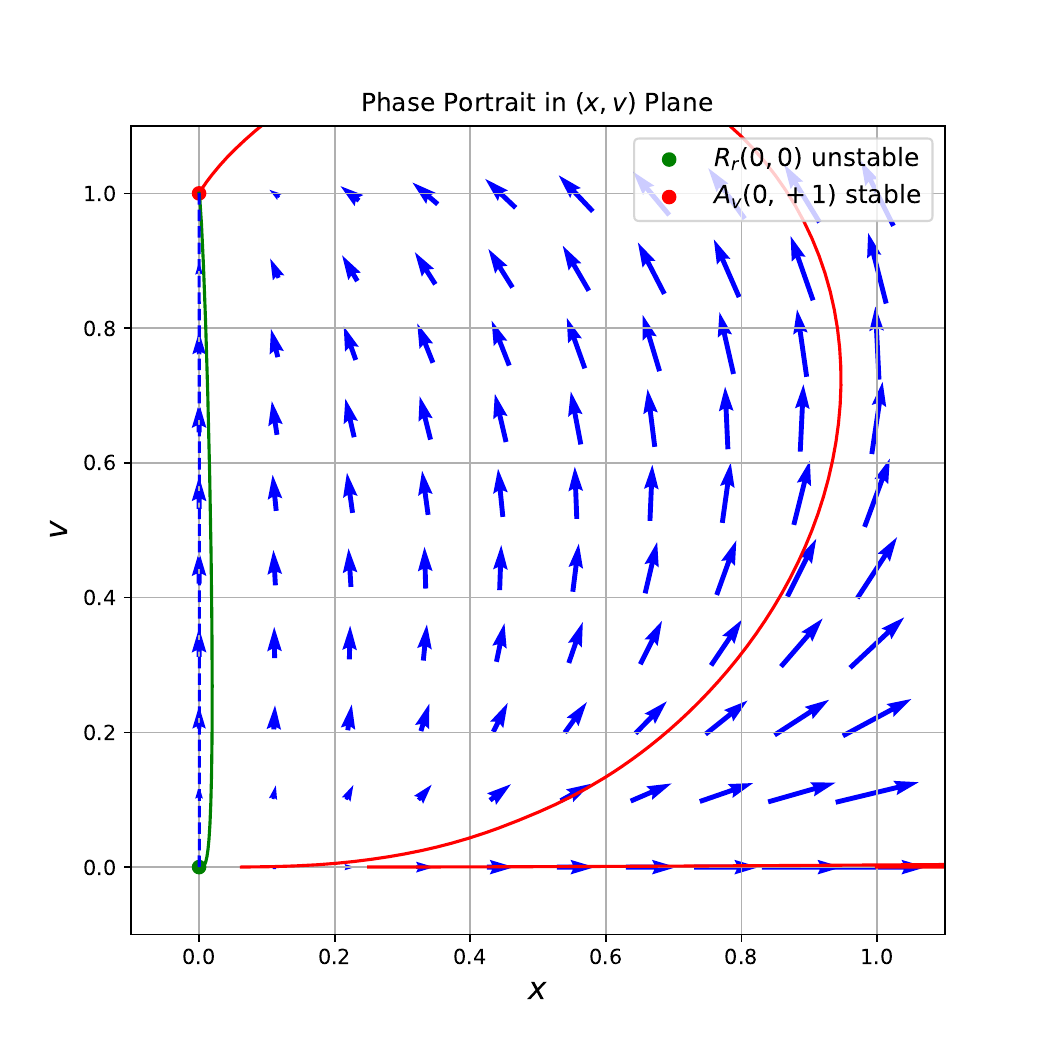}
        \label{fig:ChatGPT_scalarLCDM_7D_3D_set_of_sim_diff_eqns_phase_portraits_xy1}   
    \end{subfigure}
    \vspace{-1cm}
    \caption{ We illustrate the phase portrait of the 3D model. Left Panel: We illustrate the phase portrait of the \( (m,x) \). We cannot observe the $(m,x)=(0,0)$ as a unstable initial point, but we can observe it as a saddle point. This is due to that this phase portrait is a projection of the 3D case, where both points appear. Center Panel: We illustrate the phase portrait of the \( (m,v) \) Right Panel: We illustrate the phase portrait of the \( (x,v) \). 
    [See section \ref{sec:phase_portraits_of_the_6D_to_3D_system}].}
    \label{fig:ChatGPT_scalarLCDM_7D_3D_set_of_sim_diff_eqns_phase_portraits_mxy1}
\end{figure*}

In \refF{fig:ChatGPT_scalarLCDM_7D_3D_set_of_sim_diff_eqns_phase_portraits_mxy1}, we plot the $(m,x)$-plane phase portrait, and we find the following points:
\begin{itemize}
	\item \textit{The unstable point, $R_r(0,0)$}, is characterised by a vanishing matter energy density ratio, vanishing scalar kinetic energy density ratio, vanishing scalar potential energy density ratio, and dominant radiation energy density ratio.
	\item \textit{The unstable point, $R_m(0,1)$}, is characterised by a dominant matter energy density ratio, vanishing scalar kinetic energy density ratio, vanishing scalar potential energy density ratio, and vanishing radiation energy density ratio.	
\end{itemize}

We observe that the dynamical field starts from the repellers  $R_r(0,0)$ and $R_m(0,1)$ and they go towards the attractor $A_v(0,0)$, which is not apparent in this plot, since this is the profile at $v=0$.

The physical picture is that we start with high radiation energy density ratio, while vanishing energy density ratio for matter, kinetic and potential scalar of the universe, i.e. $m=\Omega_m = \Omega_x = \Omega_v=0$, and then we move to a region where energy density ratio for the potential term of the $V[N(t)]$-field. 

This signifies the transition 
from a 
\textit{vanishing matter energy density ratio, vanishing scalar kinetic energy density ratio, vanishing scalar potential energy density ratio, and dominant radiation energy density ratio} 
epoch, 
towards a 
\textit{vanishing matter energy density ratio, vanishing scalar kinetic energy density ratio, dominant scalar potential energy density ratio, and vanishing radiation energy density ratio}
epoch.

Or in other words, this signifies the transition 
from a 
\textit{dominant radiation energy density ratio} 
epoch, 
towards a 
\textit{dominant potential energy density ratio}
epoch.

Or in other words, this signifies the transition 
from a 
\textit{dominant radiation energy density ratio} 
epoch, 
towards a 
\textit{dominant cosmological constant energy density ratio}
epoch.

In \refF{fig:ChatGPT_scalarLCDM_7D_3D_set_of_sim_diff_eqns_phase_portraits_mxy1}, we plot the $(m,v)$-plane phase portrait, and we find the following points:

\begin{itemize}
	\item \textit{The unstable point, $R_r(0,0)$}, is characterised by a vanishing matter energy density ratio, vanishing scalar kinetic energy density ratio, vanishing scalar potential energy density ratio, and dominant radiation energy density ratio.
	\item \textit{The unstable point, $R_m(1,0)$}, is characterised by a dominant matter energy density ratio, vanishing scalar kinetic energy density ratio, vanishing scalar potential energy density ratio, and vanishing radiation energy density ratio.	
	\item \textit{The    stable point, $A_v(0,1)$}, is characterised by a vanishing matter energy density ratio, vanishing scalar kinetic energy density ratio, dominant scalar potential energy density ratio, and vanishing radiation energy density ratio.
\end{itemize}

We observe that the dynamical field starts from the repellers  $R_r(0,0)$ and $R_m(1,0)$ and they go towards the attractor $A_v(0,0)$, which is clearly apparent in this plot, since this is the profile at $x=0$.

The physical picture is that we start with high radiation energy density ratio, while vanishing energy density ratio for matter, kinetic and potential scalar of the universe, i.e. $m=\Omega_m = \Omega_x = \Omega_v=0$, and then we move to a region where energy density ratio for the potential term of the $V[N(t)]$-field. We also start from a dominant energy density ratio for matter, but vanishing kinetic and potential scalar of the universe, towards a dominant potential scalar term.

This signifies the transition 
from a 
\textit{vanishing matter energy density ratio, vanishing scalar kinetic energy density ratio, vanishing scalar potential energy density ratio, and dominant radiation energy density ratio} 
epoch
or 
\textit{dominant matter energy density ratio, vanishing scalar kinetic energy density ratio, vanishing scalar potential energy density ratio, and vanishing radiation energy density ratio} 
, 
towards a 
\textit{vanishing matter energy density ratio, vanishing scalar kinetic energy density ratio, dominant scalar potential energy density ratio, and vanishing radiation energy density ratio}
epoch.

Or in other words, this signifies the transition 
from a 
\textit{dominant radiation energy density ratio} 
epoch, 
or a \textit{dominant matter energy density ratio} 
epoch, 
towards a 
\textit{dominant potential energy density ratio}
epoch.

Or in other words, this signifies the transition 
from a 
\textit{dominant radiation energy density ratio} 
epoch, 
or a \textit{dominant matter energy density ratio} 
epoch, 
towards a 
\textit{dominant cosmological constant energy density ratio}
epoch.

\textit{Note that the scalar potential energy density ratio domination epoch corresponds to a $\Lambda$ energy density ratio domination epoch in the far future.}


In \refF{fig:ChatGPT_scalarLCDM_7D_3D_set_of_sim_diff_eqns_phase_portraits_mxy1}, we plot the $(x,v)$-plane phase portrait, and we find the following points:

\begin{itemize}
	\item \textit{The unstable point, $R_r(0,0)$}, is characterised by a vanishing matter energy density ratio, vanishing scalar kinetic energy density ratio, vanishing scalar potential energy density ratio, and dominant radiation energy density ratio.
	\item \textit{The    stable point, $A_v(0,1)$}, is characterised by a vanishing matter energy density ratio, vanishing scalar kinetic energy density ratio, dominant scalar potential energy density ratio, and vanishing radiation energy density ratio.
\end{itemize}

We observe that the dynamical field starts from the repellers  $R_r(0,0)$ and they go towards the attractor $A_v(0,0)$, which is clearly apparent in this plot, since this is the profile at $\Omega_m=0$.

The physical picture is that we start with vanishing radiation energy density ratio, while vanishing energy density ratio for matter, kinetic and potential scalar of the universe, i.e. $m=\Omega_m = \Omega_x = \Omega_v=0$, and then we move to a region where energy density ratio for the potential term of the $V[N(t)]$-field. We also start from a low scalar kinetic term energy density ratio, but vanishing potential scalar of the universe, towards a dominant scalar potential  term.

This signifies the transition 
from a 
\textit{vanishing matter energy density ratio, vanishing scalar kinetic energy density ratio, vanishing scalar potential energy density ratio, and dominant radiation energy density ratio} 
epoch
or 
\textit{vanishing matter energy density ratio, low scalar kinetic energy density ratio, vanishing scalar potential energy density ratio, and vanishing radiation energy density ratio} 
, 
towards a 
\textit{vanishing matter energy density ratio, vanishing scalar kinetic energy density ratio, dominant scalar potential energy density ratio, and vanishing radiation energy density ratio}
epoch.

Or in other words, this signifies the transition 
from a 
\textit{dominant radiation energy density ratio} 
epoch, 
or a \textit{low scalar kinetic term energy density ratio} 
epoch, 
towards a 
\textit{dominant potential energy density ratio}
epoch.

Or in other words, this signifies the transition 
from a 
\textit{dominant radiation energy density ratio} 
epoch, 
or a \textit{low scalar kinetic term energy density ratio} 
epoch, 
towards a 
\textit{dominant cosmological constant energy density ratio}
epoch.

\textit{Note that the scalar potential energy density ratio domination epoch corresponds to a $\Lambda$ energy density ratio domination epoch in the far future.}

\section{Numerical solution of $\Lambda$CDM and $\phi\Lambda$CDM model comparison}\label{sec:numerical_solutions_comparison_of_models}

We perform a comparison between the $\Lambda$CDM model (DA investigated in \refApp{sec:wLCDMcosmology_phaseportrait}) and the $\phi\Lambda$CDM model (DA investigated in \refS{sec:Dynamical_Analysis_phiLambdaCDM}).

We identify similar numerical solutions of the two models, which is an expected results for a simple modification of gravity, such as this additional dynamical scalar fields.
In agreement with current observation, we find similar behaviour for different epochs, at the level of the mean values,
\begin{itemize}
\item matter-radiation equality epoch redshift is the same \( z^{mr}_{\Lambda\text{CDM}} = z^{mr}_{\phi\Lambda\text{CDM}} = 1484\),
\item radiation-dark energy equality redshift epoch is the same \( z^{r\Lambda}_{\Lambda\text{CDM}} = 6.5 \simeq z^{r\phi}_{\phi\Lambda\text{CDM}} = 6.8\), 
\item matter-dark energy equality epoch redshift is the same \( z^{m\Lambda}_{\Lambda\text{CDM}}  = z^{m\phi}_{\phi\Lambda\text{CDM}} = 0.3 \), 
\item in the far past, \( t \to -\infty \Leftrightarrow N \to -\infty \Leftrightarrow z \to -\infty\), the models have a radiation domination epoch, i.e a radiation repeller.
\item in the far future, \( t \to +\infty \Leftrightarrow N \to -\infty \Leftrightarrow z \to +\infty \), the models have cosmological constant domination epoch, i.e. a \( \Lambda \) attractor, i.e. a dark energy attractor, i.e. a de Sitter attractor.
\item in the recent past, \( N \to -4 \Leftrightarrow z \to 50\), the models have matter domination epoch, i.e. matter saddle point.
\end{itemize}

    \begin{figure*}[h!]
    \centering 
    \includegraphics[width=140mm]{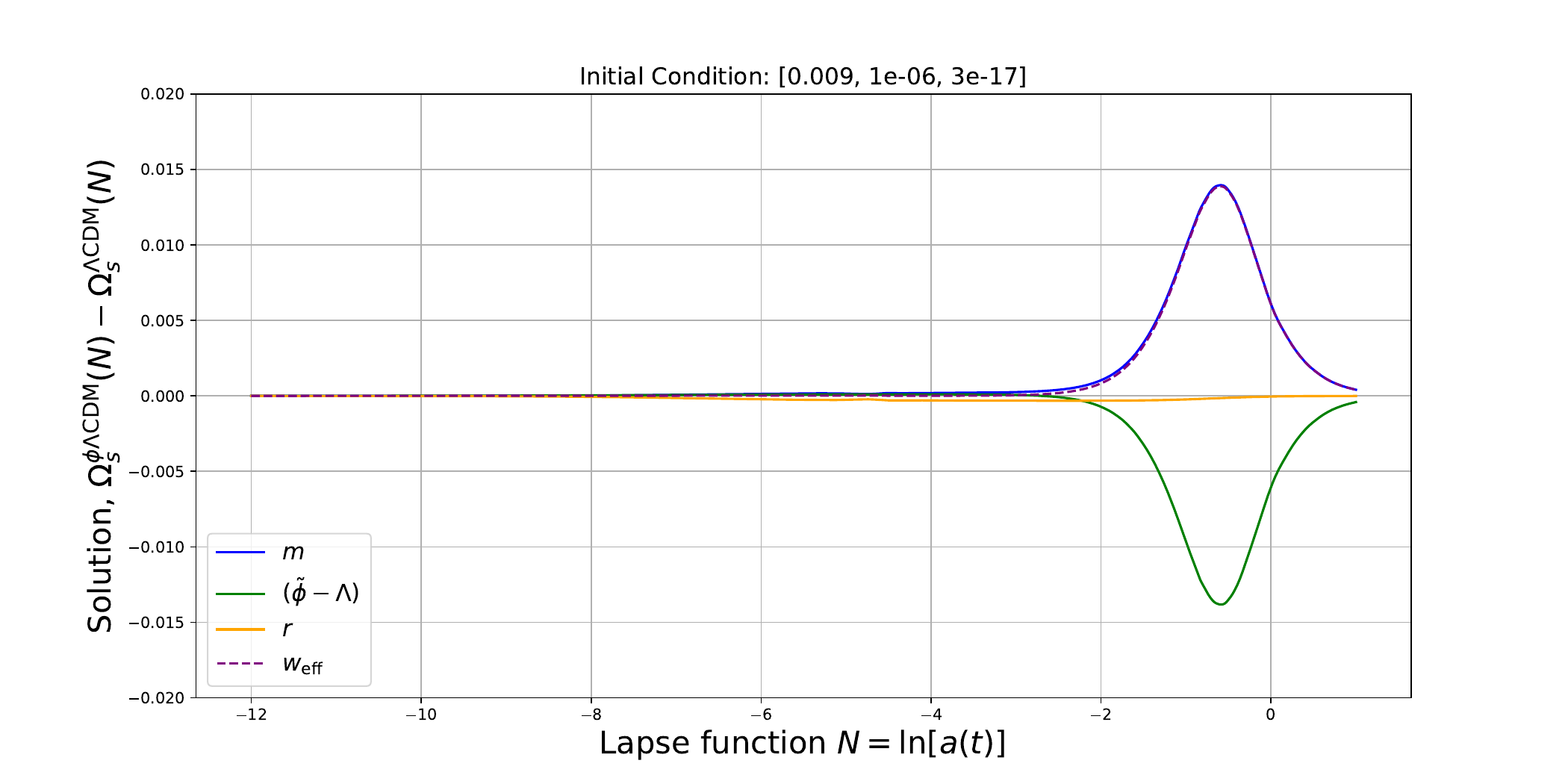} 
    \caption{\label{fig:ChatGPT_6D_to_3D_phiLCDM_comparison_with_3D_4eqns_LCDM_set_of_sim_diff_eqns_num_solutions}  Comparison of numerical solutions of the energy density ratio epochs between $\Lambda$CDM and $\phi\Lambda$CDM. [See \refS{sec:numerical_solutions_comparison_of_models}] }
    \end{figure*}

In \refF{fig:ChatGPT_6D_to_3D_phiLCDM_comparison_with_3D_4eqns_LCDM_set_of_sim_diff_eqns_num_solutions}, we present the difference of energy density ratios between the two models, as a function of lapse time, $\Omega_s^{\phi\Lambda{\rm CDM}}(N)-\Omega_s^{\Lambda{\rm CDM}}(N)$, for the different species, $s$. We find that 
\begin{itemize}
\item the radiation between the two models agrees at mosts times.  
\item the matter between the two models does not agree at mosts times, it diverges about max $1.5\%$ in favour of $\phi\Lambda$CDM, at late times.
\item the dark energy the constant and the dynamical between the two models agree at mosts times, it diverges about max $1.5\%$ in favour of $\Lambda$CDM, at late times.
\item the equation of state between the two models, agrees at mosts times, it diverges about max $1.5\%$ in favour of $\Lambda$CDM at late times.
\end{itemize}

\section{Phase portraits of $\Lambda$CDM and $\phi\Lambda$CDM model comparison}\label{sec:phasespace_criticalpoints_comparison_of_models}

Both models have the following critical points: the models have a radiation domination epoch, i.e a radiation repeller; the models have cosmological constant domination epoch, i.e. a \( \Lambda \) attractor, i.e. a de Sitter attractor; the models have matter domination epoch, i.e. matter saddle point.

The $\Lambda$CDM model has the following transitions from a radiation domination epoch to a matter domination epoch to a dark energy/cosmological constant domination epoch.

The $\phi\Lambda$CDM model has, for any scalar field, assuming any scalar potential, the model has the the following transitions:
\begin{itemize}
\item transition 
from a 
\textit{dominant radiation energy density ratio} 
epoch, 
towards a 
\textit{dominant cosmological constant energy density ratio}
epoch.
\item transition 
from a 
\textit{dominant radiation energy density ratio} 
epoch, 
or a \textit{dominant matter energy density ratio} 
epoch, 
towards a 
\textit{dominant cosmological constant energy density ratio}
epoch.
\item transition 
from a 
\textit{dominant radiation energy density ratio} 
epoch, 
or a \textit{low scalar kinetic term energy density ratio} 
epoch, 
towards a 
\textit{dominant cosmological constant energy density ratio}
epoch.
	\end{itemize}
	We find that the $\phi\Lambda$CDM model has a richer phenomenology than the standard $\Lambda$CDM model.

\section{Conclusions and discussion}\label{sec:conclusion_discussion}

In summary in this work we study the $\phi\Lambda$CDM model and the standard $\Lambda$CDM model through the scope of dynamical analysis. The $\phi\Lambda$CDM model is more complete, richer and simpler than the one that we found in the literature. In particular, we take into account all the so far discovered epochs, radiation, matter, and dark energy epochs. We also incorporate a continuity equation that allows simpler dynamics, without the consideration of a scalar potential, and the form of the field. In contrast in the literature there is no study including the radiation epoch for this model.

The relevance of the \(\phi\Lambda\)CDM model lies in its ability to provide a dynamical interpretation of the cosmological constant while remaining observationally consistent with \(\Lambda\)CDM. Unlike the standard model, where \(\Lambda\) is a fixed parameter, this framework introduces a free parameter that governs the interaction between the scalar field and the effective cosmological constant. While the late-time attractor remains a de Sitter phase, ensuring agreement with observations, variations in this parameter influence the transition dynamics between cosmic epochs. This allows for a richer phase-space structure and may provide insights into early-universe behavior or deviations from standard \(\Lambda\)CDM predictions in non-standard scenarios. 

We find that both models have the following critical points: a radiation domination epoch, i.e a radiation repeller; cosmological constant domination epoch, i.e. a \( \Lambda \) attractor, i.e. a de Sitter attractor; and a matter domination epoch, i.e. matter saddle point.

The $\Lambda$CDM model has the transition from a radiation domination epoch to a matter domination epoch to a dark energy/cosmological constant domination epoch.

The $\phi\Lambda$CDM model has the previous transition as well as several new transition that have not been discussed in the literature. In particular we find the main new transition
from a 
\textit{dominant radiation energy density ratio} 
epoch, 
or a \textit{low scalar kinetic term energy density ratio} 
epoch, 
towards a 
\textit{dominant cosmological constant energy density ratio}
epoch.

Future work will involve investigating the dynamics of various potentials, as well as exploring more sophisticated and advanced systems, such as $f(R)$ gravity, Horndeski theories, non-Riemannian cosmologies, and functor of actions theories frameworks \cite{2023FoPh...53...29N,Ntelis:2024fzh}.


\section*{ACKNOWLEDGEMENTS}

	We acknowledge open libraries support \texttt{IPython} \cite{4160251}, \texttt{Matplotlib} \cite{Hunter:2007}, \texttt{NUMPY} \cite{Walt:2011:NAS:1957373.1957466} \texttt{SciPy} \cite{2019arXiv190710121V}, 



\appendix

\section{Triples in different epochs}

In \refT{tab:epoch_time_triplet_different_energy_density_ratio_species_equivalences}, we present the distinct epochs associated with the time triplet corresponding to the various species' energy density ratio equivalences.

\begin{table}[h!]
\centering
\begin{tabular}{|c|c|c|c|}
\hline
\textbf{Epoch} & lapse function, $N(t)$ & redshift, $z(t)$ & scale factor, $a(t)$ \\ \hline
today, $0$ & 0 & 0 & 1 \\ \hline
matter-Dark energy, $m \Lambda$ & -0.282 & 0.326 & 0.753 \\ \hline
radiation-Dark energy, $r \Lambda$ & -2.04 & 7.69 & 0.130 \\ \hline
decoupling & -7.01 & 1089 & $9 \times 10^{-4}$ \\ \hline
recombination & -7.01 & 1100 & $9 \times 10^{-4}$ \\ \hline
matter-radiation, $m r$ & -7.31 & 1500 & $7 \times 10^{-4}$ \\ \hline
initial, $i$ & -12 & $2 \times 10^5$ & $6 \times 10^{-6}$ \\ \hline
\end{tabular}
\caption{\label{tab:epoch_time_triplet_different_energy_density_ratio_species_equivalences} Classification of time triplet epoch according to species energy density ratio.}
\end{table}

\section{Dynamical system analysis on $\Lambda$CDM}
\label{sec:wLCDMcosmology_phaseportrait}

Following \citet{Bahamonde:2017ize}, we model the $\Lambda$CDM-Cosmology, and we apply the dynamical system analysis as follows.

\subsection{The $\Lambda$CDM model}
We know that assuming a non-curved, flat FLRW metric, and standard General Relativity Gravity, namely, $R$-Gravity, we obtain the following cosmological scenario, using a cosmological constant and cold dark matter, and assuming the least action principle, the $\Lambda$CDM-Cosmology is governed by the background Einstein Field Equations equations, 
$
	G_{\mu\nu} = \kappa^2 T_{\mu\nu} \; , 
$ where $\kappa^2=\frac{8\pi G_{\rm N}}{c^4}$.
These result to Friedman equations which have the species matter, radiation, and cosmological constant, $\Lambda$.
and we also use the continuity equation
${T^{\mu \nu}}_{;\mu}=0 \; $ .
For each different energy density species we have a different equation of state:
$
	\{ w_m, w_r, w_\Lambda \}= \{ 0, 1/3, -1 \} \; ,
$
which corresponds to a different energy density species continuity equations. 

\subsection{Defining the dimensionless variables for $\Lambda$CDM}
To apply the dynamical analysis, we need to simplify mathematically the problem, and we define the following dimensionless variables
\begin{eqnarray}
x(t) = \Omega_m(t) =  \frac{\kappa^2 \rho_{\rm m}(t)}{3H^2(t)}, \quad
y(t) = \Omega_r(t) =  \frac{\kappa^2 \rho_{\rm r}(t)}{3H^2(t)}, \quad
z(t) = \Omega_\Lambda(t) =  \frac{\kappa^2 \rho_{\rm \Lambda}}{3H^2(t)} \equiv   \frac{\Lambda}{3H^2(t)} \; .
\end{eqnarray}

Using the previous definitions, we write the 3D set of differential equations  simply
\begin{eqnarray}
	\label{eq:wLCDM_system_of_simultaneous_differential_equations_of_xyz_4eqns_with_1stFriedman_as_a_function_of_conformal_time_eta_1}
	x' &=& x( -3 + 3x + 4y)\\
	\label{eq:wLCDM_system_of_simultaneous_differential_equations_of_xyz_4eqns_with_1stFriedman_as_a_function_of_conformal_time_eta_2}
	y' &=&  y( -4 + 3x + 4y)\\
	\label{eq:wLCDM_system_of_simultaneous_differential_equations_of_xyz_4eqns_with_1stFriedman_as_a_function_of_conformal_time_eta_3}
	z' &=& + (1-x-y)  (3 x + 4 y )
\end{eqnarray}
where we integrate in the previous set of equations; the 1st Friedmann equation for the model:
$
	z = 1 - x - y \; .
$

	According to these definition, the effective equation of state, for this model, is given by
$
	w_{\rm eff} 
	= \sum_{s \in \left\{ m, r, \Lambda \right\}} w_s(t) \Omega_s(t) = \frac{1}{3} y(t) - z(t)  \; .
$

\subsection{$\Lambda$CDM model: system in 2D, 3D projected to 2D and 3D approaches}\label{sec:LambdaCDM_2D_to_3D}
We solve the system analytically and numerically in 2D, 3D projected to 2D and in 3D and we find similar results for the epoch evolution of the different species in our $\Lambda$CDM model. We also apply the DA to the system and we obtain similar phase portraits and results, in 2D in 3D projected to 2D and in 3D. The 3D projected to 2D and the 3D cases, are not customary done in the literature, however we find that they are equivalent. In this work, we provide 3D equivalent approach. We use the solutions and the DA for the 3D model.

\subsection{Dynamical analysis of the 3D set of differential equations of $\Lambda$CDM model}
\label{sec:wLCDMcosmology_3D_xyz_trjectories}

\paragraph{}We solve the system of simultaneous differential equations of $(x,y,z)$ as a function of lapse function $N$, described by equations \ref{eq:wLCDM_system_of_simultaneous_differential_equations_of_xyz_4eqns_with_1stFriedman_as_a_function_of_conformal_time_eta_1}-\ref{eq:wLCDM_system_of_simultaneous_differential_equations_of_xyz_4eqns_with_1stFriedman_as_a_function_of_conformal_time_eta_3}, 
numerically, using the $\texttt{ipython}$ library $\texttt{scipy.integrate.solveivp}$. The selection of the 3D system is explained in \refApp{sec:LambdaCDM_2D_to_3D}

We use different initial conditions. However, the most physical one, is the one where 
\begin{eqnarray}
	\left\{ x(N_i), y(N_i), z(N_i) \right\} = \{0.01,0.99,0\} \; . 	
\end{eqnarray}
This is the most physical one, since the sum of all these functions should be equal to 1 at all times,
and also because we have good evidence that initial the universe was mostly field with radiation, and some dark energy, and some matter energy densities, although the latter two should be vanishing. Note that the range for the lapse function is $\Delta N = [-1,1]$.

We find that the system with initial conditions for which initial dark energy density ratio, is higher than or comparable with the initial total matter density ratio, 
 then the total matter density ratio, is not allowed by the system to evolve, and it is quickly suppressed. While in the case for which initial dark energy density ratio, is lower than the initial total matter density ratio, 
 then the total matter density ratio, is allowed by the system to evolve better, and it is quickly increases to describe the matter dominated epoch, as we know so far by observations and previous theories.
We find the numerical solutions of the 3D system.

\subsubsection{Stability analysis on 3D system of $\Lambda$CDM}\label{sec:LCDMcosmology_phaseportrait}
To analyze the stability of the critical points, we need to compute the Jacobian matrix of the system. The Jacobian matrix \(J\) is given by the matrix of partial derivatives of the right-hand sides of the system with respect to the variables \(x\), \(y\), and \(z\).
We find its corresponding eigenvalues to characterise the critical points.

    \begin{figure*}[ht!]
    \centering 
    \includegraphics[width=80mm]{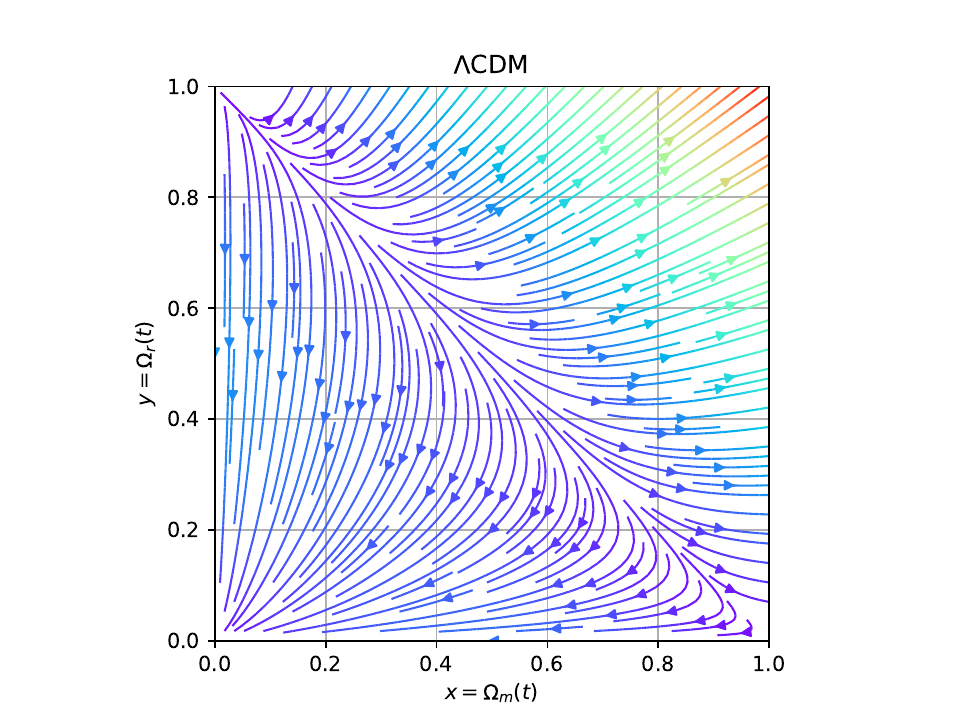} 
    \caption{\label{fig:LCDMcosmology_phaseportrait}   We illustrate the phase portrait of $\Lambda$CDM cosmology. [See \refApp{sec:LCDMcosmology_phaseportrait}] }
    \end{figure*}

The summary of stability is the following:

- The critical point \( (x, y) = (0, 0) \) is a stable point. The eigenvalues of this matrix are \( -3 \), \( -4 \), and \( 0 \). Since two eigenvalues are negative and one is zero, this critical point is a stable point (Attractor).

- The critical point \( (x, y) = (0, 1) \) is unstable. The eigenvalues of this matrix are \( 1 \), \( 4 \), and \( 0 \). Since two eigenvalues are positive and one is zero, this critical point is unstable (repelling).

- The critical point \( (x, y) = (1, 0) \) is a saddle point. The eigenvalues of this matrix are \( 6 \), \( -1 \), and \( 0 \). Since one eigenvalue is positive and one is negative, this critical point is also a saddle point.

This provides insight into the dynamics of your system near these points. 

We present the results in \refF{fig:LCDMcosmology_phaseportrait}.

We find that the universe starts with a radiation energy density ratio domination epoch at the unstable repeller point $R^{xy}(0,1) = R^{mr}(0,1)$, transits to a matter energy density ratio domination epoch, at the saddle point, $R^{xy}(1,0) = R^{mr}(1,0)$, and it is attracted in the far future to the stable attractor point, $R^{xy}(0,0) = R^{mr}(0,0)$, i.e it results to a dark energy density ratio domination epoch.

In simple terms we conclude the following. 

We find that the universe starts with a radiation energy density ratio domination epoch, then it transits to a matter energy density ratio domination epoch and it is attracted, in the far future, to a dark energy density ratio domination epoch.

\section{Details of the $\phi \Lambda$CDM dynamical analysis}
Below, we provide all the necessary details of the $\phi \Lambda$CDM dynamical analysis.

\subsection{The $\phi\Lambda$CDM action}

The $\phi\Lambda$CDM action is written as 
\begin{eqnarray}
	S_{\phi\Lambda\text{CDM}} = c^3 \int d(ct) d^3x \sqrt{-g} \left[ \frac{1}{16 \pi G_{\rm N}} (R) + \mathcal{L}_{\rm mr} - \frac{1}{2} g_{\mu\nu}\partial^{\mu}\phi \partial^{\nu}\phi + V[t, \Lambda; \phi(t)]  \right]
\end{eqnarray}
where $c$ is the sped of light, $G_{\rm N}$ is the Newton constant, $g$ is the determinant of the metric, $g_{\mu\nu}$, $R=R[g]$ is the ricci scalar, $\Lambda$ is the cosmological constant, $\mathcal{L}_{\rm mr}[g,\psi_{mr}]$ is the lagrangian describing matter and radiation field, $\psi_{\rm mr}$, $\phi$ is a scalar dynamical field, with a kinetic term, described by partial derivatives, and $V[t, \Lambda; \phi(t)]$ is its potential. Assuming the least actionic principle, varying the aforementioned action, we are lead to the following equations of motion of the system.

\subsection{The cosmological constant, $\Lambda$, in $\phi \Lambda$CDM }\label{sec:description_of_Lambda_in_phiLambdaCDM_model}
In the standard cosmological model, $\Lambda$ cosmological constant describe dark energy.
In our case, the $\phi$CDM model, now we have a dynamical scalar field which describes dark energy, however the cosmological constant is still inherent in our model,
through the following consideration.

The choice of the potential of this scalar field is 
\[
V(\phi) = V[t, \Lambda; \phi(t)]= V_0(\Lambda) e^{- \kappa \lambda \phi(t)}		
\] 
where $V_0(\Lambda)$ is the normailsation constant of the potential.

To retrieve the $\Lambda$CDM model from $\phi$CDM model,
we need the condition: 
\[
V_0(\Lambda) = - \frac{1}{16 \pi G_{\rm N}} 2 \Lambda \hspace{1cm} \text{and} \hspace{1cm} 	\phi(t) = 0	
\] 
In that case the potential becomes 
\[
V(\phi) = - \frac{1}{16 \pi G_{\rm N}} 2 \Lambda
\] 
and the action model 
\begin{eqnarray}
	S_{\phi\Lambda\text{CDM}} = c^3 \int d(ct) d^3x \sqrt{-g} \left[ \frac{1}{16 \pi G_{\rm N}} R + \mathcal{L}_{\rm mr} - \frac{1}{2} g_{\mu\nu}\partial^{\mu}\phi \partial^{\nu}\phi + V[\Lambda; \phi(\Lambda)]  \right]
\end{eqnarray}
becomes
\begin{eqnarray}
	S_{\Lambda\text{CDM}} = c^3 \int d(ct) d^3x \sqrt{-g} \left[ \frac{1}{16 \pi G_{\rm N}} R + \mathcal{L}_{\rm mr} - \frac{1}{16 \pi G_{\rm N}} 2 \Lambda  \right] \: .
\end{eqnarray}

\label{Bibliography}


\bibliographystyle{unsrtnat_arxiv} 

\bibliography{Bibliography} 

\end{document}